# FLAS: a combination of proactive and reactive auto-scaling architecture for distributed services

Víctor Rampérez[a,∗], Javier Soriano[a], David Lizcano[b], Juan A. Lara[b]

[a]*Universidad Politécnica de Madrid (UPM), 28660 - Boadilla del Monte, Madrid, Spain*
[b]*Madrid Open University (UDIMA), 28400 Collado Villalba, Madrid, Spain*

**Abstract**

Cloud computing has established itself as the support for the vast majority of emerging technologies, mainly due to the characteristic of elasticity it offers. Auto-scalers are the systems that enable this elasticity by acquiring and releasing resources on demand to ensure an agreed service level. In this article we present FLAS (**F**orecasted **L**oad **A**uto-**S**caling), an auto-scaler for distributed services that combines the advantages of proactive and reactive approaches according to the situation to decide the optimal scaling actions in every moment. The main novelties introduced by FLAS are (i) a predictive model of the high-level metrics trend which allows to anticipate changes in the relevant SLA parameters (e.g. performance metrics such as response time or throughput) and (ii) a reactive contingency system based on the estimation of high-level metrics from resource use metrics, reducing the necessary instrumentation (less invasive) and allowing it to be adapted agnostically to different applications. We provide a FLAS implementation for the use case of a content-based publish-subscribe middleware (E-SilboPS) that is the cornerstone of an event-driven architecture. To the best of our knowledge, this is the first auto-scaling system for content-based publish-subscribe distributed systems (although it is generic enough to fit any distributed service). Through an evaluation based on several test cases recreating not only the expected contexts of use, but also the worst possible scenarios (following the Boundary-Value Analysis or BVA test methodology), we have validated our approach and demonstrated the effectiveness of our solution by ensuring compliance with performance requirements over 99% of the time.

## 1. Introduction

We have seen how in just a few years society has transformed and evolved towards an increasingly digitalized world, where all aspects of its daily life depend on technology and more specifically on Internet services. For all these reasons, it is not surprising that computer resources are now an essential utility in modern societies on a par with electricity, gas or water. As a result, cloud computing emerged as a way to provide computing resources as a service, that is, on-demand computing resources that users acquire on a pay-as-you-go basis.

Cloud computing has been consolidated as a support for the vast majority of current and emerging technologies. For example, the widespread adoption of event-driven architectures [1], which are essential for real-time-sensitive digital business such as IoT (Internet of Things), has been possible because cloud computing is able to provide an infrastructure that meets the requirements demanded by high-performance distributed systems (publish/subscribe message brokers, distributed stream processing systems or distributed datastores), which are the cornerstones of these architectures [2, 3, 4].

The key feature of Cloud Computing is elasticity, which is the capability to acquire and release resources on demand to meet end-user requirements, which are formally expressed through Service Level Agreements or SLAs. However, it is not a trivial task to decide the exact amount of resources needed at any given time to meet these SLAs. There are several types of SLAs depending on the magnitude that end users want to manage such as performance, cost or energy consumption. Therefore, an auto-scaling system is desirable to free the users from the burden of adjusting allocated resources to meet SLAs at any given time. The main objective of auto-scaling systems is to avoid both over-provisioning and under-provisioning of resources, which would increase the cost and violate the SLA respectively.

Many auto-scaling systems have been developed in both the literature and the industry proposing different approaches to the problem. These auto-scaling techniques are classified into two major groups: (i) reactive techniques, where the scaling action is in reaction to a change in the system,

∗Corresponding author
*Email addresses:* `v.ramperez@upm.es` (Víctor Rampérez ), `javier.soriano@upm.es` (Javier Soriano), `david.lizcano@udima.es` (David Lizcano), `juanalfonso.lara@udima.es` (Juan A. Lara)

and therefore does not anticipate such a change; and (ii) predictive or proactive techniques, which attempt to anticipate future changes in the system by performing the necessary scaling actions before such changes occur [5].

A scaling action is defined by the specific values of its dimensions, i.e. which resource is to be scaled (CPU, memory, network, etc), when to scale, how many resources are to be added or removed, and how to scale (horizontal or vertical scaling). An auto-scaling system can be seen as a system that returns a specific scaling action (with specific values to each of the dimensions) based on a series of parameters or input information provided to it (e.g. SLA, workload, application information to be scaled, predictive models, threshold-based scaling rules, etc) to ensure compliance with a SLA. Because of this, auto-scaling systems are quite complex and existing approaches usually focus only on one type of SLA (e.g. performance, cost or energy consumption), one or two dimensions of the scaling actions (e.g. when to scale and how much) and a specific application or type of application (e.g. distributed stream processing systems).

In order to achieve the desired elasticity of an application, several works and authors highlight the need to understand the relationship between the low-level behavior of that application and the high-level parameters of the SLA to be ensured [6, 7, 8, 9, 10, 11]. Therefore, auto-scale systems would have to be equipped with the necessary mechanisms that allow them to establish a relationship between the low level behavior (i.e. at resource level expressed through resource metrics such as CPU usage, memory usage, etc) and the high level behavior (i.e. SLA parameters of performance, cost, etc) of the application in order to take the appropriate scaling actions to ensure compliance with the corresponding SLA. However, few works address this problem, as they take for granted the resource that is the bottleneck and therefore the resource to be scaled. For example, many jobs take for granted that the limiting resource or KPI (Key Performance Indicator) is the processing capacity and therefore their scaling action consists in increasing the processing capacity by adding more processors directly (scale-up) or more virtual machines (scale-out). We do not doubt that in these works the resource that they scale is the adequate one, since they usually demonstrate it empirically, but we defend that the study of this relation between low-level metrics and high-level metrics allows to characterize the system in a more precise way. In fact, there are several research works that point in this direction to improve their approaches in their future work [6, 12]. For example, although the resource to be scaled is processing capacity, the percentage of CPU usage may not be the most informative metric, and context changes, interruptions, or the percentage of time processors spend in user or kernel space may be more useful.

The inclusion of the information of this relationship between low and high-level metrics in an auto-scaling system considerably extends the range of applications to which such auto-scaling system can be applied, since it allows the detection of the resource that is the bottleneck and therefore the resource to be scaled regardless of the type of application and in a totally transparent way for the end user.

With all this, although there are several jobs related to auto-scaling systems in the Cloud, we have identified the following unmet needs. On the one hand, there is a need to be able to relate or establish a mapping between resource utilization metrics or low-level metrics with relevant high-level metrics in SLAs through some predictive model. This would allow identifying the resources that act as bottlenecks (KPIs) automatically (without having to assume anything), which would have to be monitored and scaled to avoid a possible SLA violation in the future or reduce unnecessary costs. On the other hand, there is a need to develop a predictive model capable of determining how fast a SLA violation situation or unnecessary over-provisioning situation can be reached in order to perform the necessary scaling action at the most convenient time, as opposed to current scenarios that only predict future workload and not how this will affect SLA compliance.

In this paper we propose FLAS (Forecasted Load Auto-Scaling) a proactive and reactive auto-scale architecture of distributed systems. FLAS works by learning and predicting patterns in the performance behavior of distributed systems in order to take the appropriate scaling decisions at any given time to ensure compliance with SLAs. The main contribution of this work, and especially of FLAS, is oriented to cover the needs previously identified and are the following: (i) a predictive model of the high-level metrics trend which allows to anticipate changes in the relevant SLA parameters (e.g. performance metrics such as response time or throughput) and (ii) a reactive contingency system based on the estimation of high-level metrics from resource use metrics, reducing the necessary instrumentation (less invasive) and allowing it to be adapted agnostically to different applications.

Due to the great importance of event-driven architectures in current technologies like IoT [1], we wanted to evaluate our auto-scaling system with a high performance distributed system like a publish-subscribe middleware. Among all the publish-subscribe systems, we have opted for content-based systems (CBPS) due to the greater complexity of their scaling actions as a result of the distribution of their internal state. More specifically, being FLAS a generic solution, we have chosen to apply it to the E-SilboPS due to the great challenge that it represented, being a CBPS that supports transparent, publisher-wise dynamic state repartitioning without client disconnection and with minimal notification delivery interruption for subscribers [10].

Due to privacy issues and commercial interests in releasing user information, there is a great lack of publicly available and realistic workloads for research and evaluation of content-based publish-subscribe systems [13]. Therefore, the evaluation has been done with synthetic work-loads through several test cases recreating not only the expected contexts of use, but other test cases representing the worst possible scenarios (following the Boundary-Value Analysis or BVA test methodology). The results of this evaluation show how the integration of proactive techniques with models to predict workload behavior, scaling time and relationship between low and high-level metrics, together with a reactive contingency system, results in a minimum violation of the established SLAs (less than 1% of run time).

The rest of the document is organized as follows: Section 2 reviews the related work analyzing the different academic and commercial solutions proposed. Section 3 introduces the system modeling and presents the problem to be addressed. The architecture of FLAS is explained in detail in Section 4. Sections 5 and 6 describe the evaluation of FLAS with a distributed content-based publish-subscribe system (E-SilboPS) through multiple test cases and analyze the quantitative results of such evaluation, respectively. Finally, the conclusions of this work are raised in Section 7 and future lines are expressed in Section 8.

## 2. Related work

Many auto-scaling systems, both academic and commercial, have been proposed recently due to the ubiquity of cloud computing and the improvement of predictive systems in recent years, using diverse approaches based on both reactive and predictive (also known as proactive) strategies [5, 14, 15, 16]. The reactive approach, widely studied in the past, is usually based mainly on threshold-based rules techniques with different variations to solve or mitigate some of the intrinsic problems of this approach, such as the use of cool-down times (also called inertia or calm) or dynamic thresholds. In recent years, more focus has been placed on predictive solutions using machine learning, reinforcement learning, queuing theory, control theory or time series analysis techniques, among others.

The vast majority of these works tend to focus on the temporal dimension (when to scale) and the quantitative dimension (how much to scale), making it obvious which resource to scale. Usually this dimension of scaling is not analyzed because it is considered trivial as a result of previous knowledge of the application to be scaled. Moreover, usually this resource is the processing capacity, taking as Key Performance Indicator (KPI) the percentage of CPU usage or some similar metric in this sense [17, 12, 16, 18, 19]. Nevertheless, there are many works that in their future lines mark the need to study other different scaling metrics, including some of the authors that highlight the limitation of the previous approach [17, 12]. For example, in [17] Lombardi, F. et al. consider the CPU as the KPI since it is the most prominent bottleneck for the type of application they scale, however they also note the intention to include memory and bandwidth in a more complete model as future work. In the same vein, the authors of [12] recognize that many resources can potentially be the bottleneck, but also focus on the CPU resource alone, justifying it as frequently being the key resource in determining performance.

Many studies have identified the need to establish some kind of relationship or mapping between the high-level metrics, in which cloud consumers are interested, and the low-level metrics offered by cloud providers, in order to establish mechanisms to ensure that the service levels demanded by cloud consumers are met. According to [7], there is a gap between monitored metrics (low-level entities) and SLAs (high-level user guarantee parameters) and none of the approaches discussed in their work deal with the mappings of low-level monitored metrics to high-level SLA guarantees necessary in cloud-like environments. In the same vein, Paschke et al. [9] highlight the problem of the poor translation of SLAs into low-level metrics, claiming that the metrics used to measure and manage performance compliance to SLA commitments are the heart of a successful agreement and that inexperience in the use and automation of performance metrics causes problems for many organizations as they attempt to formulate their SLA strategies and set the metrics needed to support those strategies. Springs et al. [8] again also clearly identify the need to address this problem, stating that a key prerequisite for meeting these goals is to understand the relationship between high-level SLA parameters (e.g., availability, throughput, response time) and low-level resource metrics, such as counters and gauges. However, it is not easy to map SLA parameters to metrics that are retrieved from managed resources. In [12] the authors claim that specifically domain experts are usually involved in translating these SLOs into lower-level policies that can then be used for design and monitoring purposes, as this often necessitates the application of domain knowledge to this problem. In [20], the authors go further and establish correlation models between absolute resource utilization metrics (i.e. "measures report about the cumulative activity counters in the operating system") and relative resource utilization metrics (i.e. "those performance measures which values are based on the data collected from the /cgroup virtual file"), demonstrating that the use of relative resource utilization metrics underestimates the capacity required and therefore are not appropriate for determining the amount of resources needed to meet performance SLA (e.g. Response Time). There are many works that make this mapping between high and low-level metrics using black-box prediction techniques (e.g. Artificial Neural Networks or ANN), which even performing very accurate predictions, do not allow to really understand the relationships between these levels of metrics [17, 11]. We believe that it is essential to be able to understand these relationships in order to characterize and classify the applications, and that is why we have opted for a statistical method based on regression that allows us to clearly interpret the established mappings with adequate predictive accuracy.

On the other hand, event-driven architectures are becoming more prevalent recently in multiple technological paradigms, with message brokers being the cornerstone of these architectures [1]. One of the best implementations of these message brokers are content-based publish-subscribe systems (CBPS) because of their ability to allow subscribers to specify their interests and only receive notifications according to those interests, as opposed to the processing overhead that subscribers to topic-based publish-subscribe systems have to perform [10, 13, 21, 22, 23]. Therefore, in this work we have opted for a content-based publish-subscribe distributed system to evaluate our implementation of FLAS. More specifically we have used our previous work, E-SilboPS [21, 22, 23], which is a content-based publish-subscribe system specifically designed to be elastic due to its scaling algorithm. Its architecture is inspired by other CBPS like SIENA [24, 25] and E-StreamHub [26]. Despite their importance, an obstacle to the research of these systems is the lack of real and publicly available workloads, due to the privacy issue involved in disclosing the interests (subscriptions) of users and other commercial interests of the companies. The authors of [13] note this problem and address it by proposing a wide-area workload generator for content-based publish-subscribe systems. For this purpose, both subscriber interests and geographic locations are generated through statistical summaries of public data traces. However, despite indicating its intention to make this generator public, it is not currently available.

F.Lombardi et al. present in [17] a work that is closely related to FLAS. In that work they introduce PASCAL, which is a predictive auto-scaling system for distributed systems by predicting workload patterns, estimating the minimum configuration required by the application and making decisions on the corresponding scaling action based on this information at each moment. More specifically, PASCAL predicts the workload input rate and estimates the application's performance at each moment in order to estimate the minimum required configuration and take the scaling decisions that will allow reaching that minimum configuration. Both FLAS and PASCAL work in two phases, a monitoring and learning phase for the generation of the predictive models and an auto-scaling phase in which the decisions about the scaling actions to be performed are made. As for the predictive part, its objective is to predict the workload input rate, while in our approach we seek to predict the trend of the relevant SLA parameters (e.g. throughput or response time), which allows us to predict how quickly a state of SLA violation can be reached. In addition, their performance estimation model currently only takes into account CPU usage, which we consider very limited, as stated above, compared to FLAS which establishes mappings between high-level metrics and a large set of low-level metrics. Furthermore, PASCAL uses models based on Artificial Neural Networks for its predictions, which although it provides them with very good prediction results, it does not allow the understanding of the relationships between these levels of metrics. Our proposal, through statistical methods based on regression, allows us to understand these relationships, and therefore detect the resource that is the bottleneck, as well as the metric(s) that monitor it (KPI) and therefore the resource to be scaled.

## 3. System model and problem statement

*3.1. System configuration model*

We consider a cluster of $M$ nodes, understanding a node as an abstract entity of infrastructure that allows to execute a software (i.e. physical or virtual machines of a cluster, containers, etc), in which the different operators of a distributed system are executed. Each operator can have several instances, and the number of instances of each operator can be increased and decreased independently. Thus, a 1-3-2 configuration indicates that there is 1 instance of operator 1, 3 instances of operator 2 and 2 instances of the third operator. Therefore, a node can be composed of a number of operator instances that can vary over time.

Through a scaling action or *sa*, the operator instances of a certain configuration can be increased or decreased. Continuing with the notation of F. Lombardi et al. in [17], a scaling action requires a time $T_{sa}$ that depends on the workload, and in case a reconfiguration of the internal state of the instances is necessary, it will also depend on the size of the state that has to be exchanged and the number of instances before and after performing the scaling action.

According to [27], a scaling action is defined by three points in time: (i) *sa* demand point ($DP_{sa}$) which is the point at which a new configuration is required, i.e. a scaling action; (ii) *sa* triggering point ($TP_{sa}$) which is the point at which a scaling action is activated, and (iii) *sa* reconfiguration point ($RP_{sa}$) is the point in time at which the scaling action has been completely terminated. Therefore, as shown in Figure 1, the time of a scaling action $T_{sa}$ is calculated as the difference between $RP_{sa}$ and $TP_{sa}$, $T_{sa} = RP_{sa} - TP_{sa}$.

In addition, a scaling action is clearly defined by specific values of four dimensions, namely:

*When.* It refers to the time at which the scaling action should be performed. As mentioned above, there are two approaches, (i) reactive techniques in which the scaling action is taken in reaction to a change (a certain condition is met) and (ii) predictive (also known as proactive) techniques that aim to anticipate changes before they occur in order to add or release the necessary resources by the time that change occurs.

*How.* It represents the type of scaling (horizontal or vertical) and the specific scaling action (scale-out/in or scale-up/down for horizontal and vertical scaling respectively).

*What.* It refers to which resource must be scaled to meet a given SLA. Some applications may be CPU bound while others may be memory bound or limited by other resources.

*How much.* It denotes the amount of resources that must be added or released to satisfy the SLA.

Although FLAS is a generic solution, for the sake of clarity in this work we are going to focus on a specific case combining predictive and reactive techniques (*when*), horizontal scaling (*how*), a CPU-bound type application, which implies that the resource to be scaled is the processing capacity (*what*) and as a first approximation we are going to add or reduce the resources to double or half in each scaling action (*how much*).

*3.2. Workload and performance model*

As indicated above, there is a wide variety of SLAs to reflect different end-user interests. In this paper we will focus on performance SLAs. In particular we will focus on the two performance metrics par excellence in this area (high-level metrics or SLA parameters) which are throughput and response time, although economic cost and energy consumption are also considered to the extent that over-provisioning is minimized. However, FLAS could work with other types of SLAs as economic cost or energy consumption by modifying the SLA parameters and providing the corresponding sources of monitoring of those parameters.

Following the modeling described in [17, 28], end users or clients interact with the distributed system by sending messages. This input rate or workload $\lambda(t)$ is defined as the number of messages received by the system per unit of time in a given instant $t$. The system has a capacity to process a certain number of messages per time unit called service time $S$. Depending on the application, the service time can be constant or can be variable depending on the system state at a time $t$, $S(t)$. The system is said to be in saturation or overload when it receives more messages per time unit than it can process in that same amount of time, i.e. $\lambda(t) > S(t)$. When the system is in saturation, the successive messages are queued, since they cannot be processed immediately. The response time $RT(t)$ is the time required by the system to process a message, which will depend on the occupation of the system, since if the system is saturated the response time will be greater because it has, in addition to the processing time $S(t)$, to wait a time in the queue before being processed. On the other hand, the throughput $X(t)$ is defined as the amount of messages that the system can process per unit of time. The behavior of both metrics will be determined by the state of the system:

Normal: input rate is less than or equal to the service rate, i.e. $\lambda(t) \leq S(t)$ and therefore the response time can be equivalent to the service rate, i.e. $RT \simeq S(t)$ and the throughput will be equal to the input rate, i.e. $X(t) \simeq \lambda(t)$ (ignoring propagation delays).

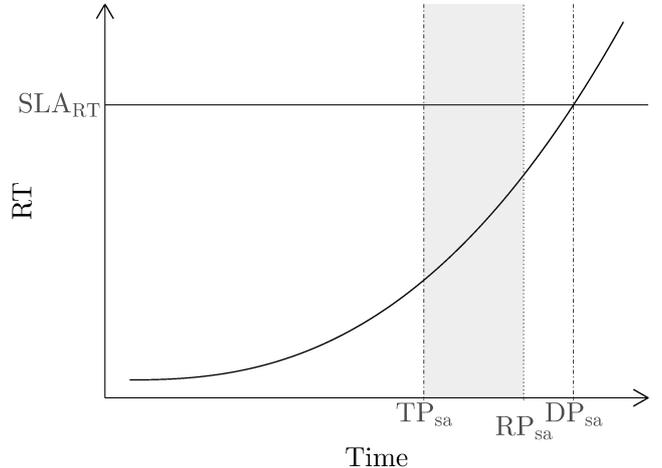

Figure 1: Example in which the response time of a distributed system increases rapidly as a consequence of saturation. To avoid exceeding the maximum response time imposed by the SLA it is necessary to perform a scale-out action that is triggered in $TP_{sa}$ and ends in $RP_{sa}$. The scaling time will be $T_{sa} = RP_{sa} - TP_{sa}$. As it is a scale-out operation that ends before the $DP_{sa}$ instant, the system will be in over-provisioning a time equal to $DP_{sa} - RP_{sa}$.

Saturation or overloaded: occurs when the input rate is higher than the service rate, i.e. $\lambda(t) > S(t)$ and therefore messages have to be queued. This causes the response time to increase exponentially (Figure 1) and the throughput to remain constant at a value close to the service rate $X(t) \simeq S(t)$.

Although the solution proposed in this paper is not based on queuing theory, we do believe that it is very interesting for the modelization of our problem. More specifically, as indicated by k.Lazowska in [28], asymptotic bound analysis provides optimistic and pessimistic limits for throughput and response time that provide rapid insights that are essential for determining the main factors affecting performance.

When the system reaches its saturation point, the system begins to act in a saturated state and performance degrades and therefore some service level objective of the SLA is often violated, especially if the trend continues. To avoid this, a scale-out action is usually performed, which allows the necessary resources to be added so that the service offered does not degrade. On the other hand, a scale-in action is necessary when the current resources are greater than those required to provide service without violating the SLA, thus saving costs or energy consumption.

Considering that the objective of the scaling actions is to return or maintain the system in a normal operating state to avoid non-compliance with SLAs, the scaling action should ideally be completed at the same time as it is demanded. Therefore, one of the objectives of the scaling systems in terms of the time dimension would be to ensure $RP_{sa} = DP_{sa}$ through reactive or proactive techniques. When this is not fulfilled, one of these two alternatives occurs:

- $RP_{sa} > DP_{sa}$: in this case the scaling action will be completed after it is needed. In the case of a scale-out it would mean an under-provisioning of resources (a configuration with less resources than required) and in the case of a scale-in it would mean an over-provisioning of resources (a configuration with more resources than required).

- $RP_{sa} < DP_{sa}$: in the opposite case to the previous one, the scaling action ends before the instant it is necessary. In this case, if the scaling action was a scale-out or a scale-in we will have over-provisioning or under-provisioning of resources respectively (Figure 1).

*3.3. Problem statement*

In general, the problem of auto-scaling is to calculate the specific scaling action (i.e. specific values at each of its dimensions) needed at each moment to ensure compliance with an SLA. Since this problem is too vast to cover its entire domain, and because elasticity is a per-application task [2], in this paper we are going to focus on a subset of applications that share a common set of characteristics. We have focused on a generic auto-scaling solution for high performance distributed systems that represents by itself a quite wide and diverse set of applications.

As mentioned above, we have focused our study on performance SLAs and the dimensions of when to scale and what resource to scale. The dimension of how to scale is application dependent and how it is designed, so it cannot be addressed in a generic solution and will have to be application specific. However, it is not a limitation of our solution as shown in the evaluation of this work. As for how much to scale, we have opted for a first approach of multiplying or dividing by two the resources to scale that allows us to verify and evaluate our solution without introducing too much complexity. However, we are working to expand our work in this direction.

More specifically, the aim is to minimize the distance between the moment when a new configuration is demanded and the moment when the scaling action covering that demand is concluded, i.e. minimize $|DP_{sa} - RP_{sa}|$, to minimize the time of over- and under-provisioning of resources. In addition, as part of the solution to this problem, we intend to develop a model that allows the mapping of low-level behavior of the application, represented by the behavior of its low-level or resource metrics, and the SLA parameters, or high-level metrics, to automatically detect which is the resource to be scaled and which low-level metric(s) are the most descriptive and useful when monitoring the system (KPIs). All of this with the main objective of not violating the SLA or minimizing the time that is being violated.

## 4. FLAS architecture

This section presents in a generic and agnostic manner the architecture of FLAS, as well as the components that compose it and its workflow. The following section (Section 5) describes the functional flow of the components that compose the architecture presented here for a specific integration with a distributed system.

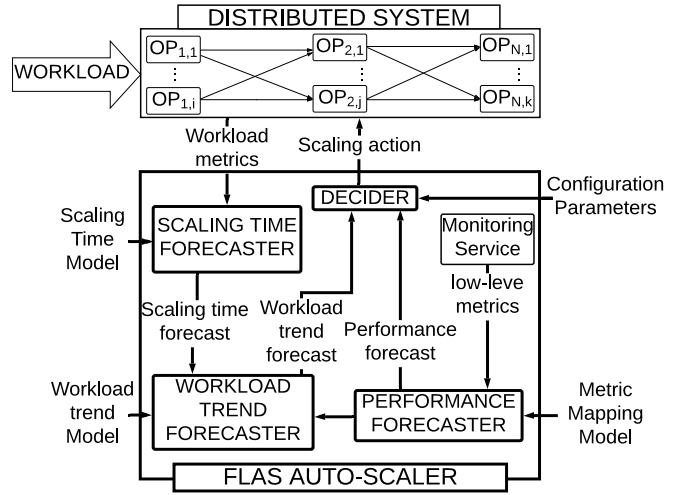

Figure 2: Functional diagram of the components that form the FLAS architecture and the integration with a distributed system.

FLAS, like other auto-scaling systems [17], works in two phases, a monitoring phase and an auto-scaling phase. In the first phase, the system collects the necessary data to acquire the knowledge needed to generate prediction models. More specifically, it collects data on the workload, the evolution of trends over time of this workload, the behavior in terms of low-level resources and the high-performance variables used in the SLA. Once these models have been generated, in the auto-scaling phase, the different modules will make use of these models to provide the necessary predictions to the decision maker, who will ultimately be responsible for deciding which auto-scaling actions should be applied, if any.

As shown in Figure 2, the FLAS architecture consists of 4 functional modules: (i) *Scaling Time Forecaster*, (ii) *Workload Trend Forecaster*, (iii) *Performance Forecaster* and (iv) *Decider*. In the following subsections these functional modules will be explained in more detail, however, the implementation of each of them may require slight adjustments to integrate with the different existing distributed systems. An example of this integration is explained in more detail in sections 5 and 6. In addition, each of these modules works as a black-box within the FLAS architecture, which allows replacing the implementation of each of these modules in a transparent manner as long as they respect the definition (interface) of these modules.

*4.1. Scaling Time Forecaster*

This module is in charge of predicting the time that a scaling action will take depending on the workload of the distributed system, i.e. $T'_{sa}(workload)$, being $T'_{sa}$ the prediction of the scaling time and $T_{sa}$ the actual time of that scaling action. Although the scaling time is mostly influenced by the workload, the predictions can be more accurate if other variables are included on which this scaling time could depend on the target distributed system. For example, in some distributed systems it is likely that the configuration before and after the scaling operation is a factor that influences this time, especially if the operators that are scaled are operators with internal state that has to be reconfigured.

*4.2. Workload Trend Forecaster*

This module is responsible for predicting the trend of system performance (i.e. SLA parameters) in the near future. Since a single prediction in a single future instant can be misleading, it is not only predicted in a single future instant, but in a time horizon $h$ (also called a forecast window) composed of several consecutive future instants. This prediction time horizon ($h$) is used to see how these predictions will evolve at different time points in the future so that the trend can be consolidated and shown in a way that is not misleading. However, the choice of this $h$-value must be made carefully as if it is too short many fluctuations may be observed, and if it is too wide relevant details of the behavior may be lost. The future instant from which these $h$ predictions are made (indicating the value of $h$ the number of consecutive predictions in time to be made, or window size) is usually calculated by means of the scaling time predicted by the *Scaling Time Forecaster* module. For example, if $h = 4$, $t_0$ is the current instant and $T'_{sa}(t_0)$ is the predicted scaling time for the current instant, then the *Workload Trend Forecaster* will make predictions in the future instants $t_i = t_0 + T'_{sa}(t_0) + i$, $\forall i \in \{0, \ldots h-1\}$

The objective of this module is to provide the *Decider* module with information on the trend of the performance of the distributed system in the prediction time horizon, indicating, for example, if during the prediction time horizon the response time will increase exponentially or if the throughput will remain constant in the next instants, both being symptoms that the system is heading towards its saturation point.

To accomplish this, during the profiling phase, the behavior of the performance metrics or SLA parameters must be recorded as a function of time to generate a time series model based on which to make these predictions in the auto-scaling phase.

*4.3. Performance Forecaster*

This module is responsible for the prediction of the performance metrics (high-level metrics) that compose the SLA. More specifically, it aims to monitor the use of low-level resources to extract a model capable of capturing the relationships between low-level and high-level metrics. This is essential because it allows to know the key performance indicators (KPIs) of the application, which indicate the bottleneck resource of the application, since if these resources are saturated (there are no more available), they lead the application to enter a state of saturation and therefore a service degradation. This knowledge is essential, because it indicates (i) the resource(s) to be monitored and by means of which metrics, and (ii) the resource to be scaled so that the application is not saturated, and therefore to avoid SLA violation. As mentioned, there are few solutions that address this problem in the literature, since they assume the resource to be scaled, and therefore the metric to be monitored, but this not only restricts the range of application of these solutions, but also, a model that reflects the behavior at low level could show that it is more effective to monitor other metric(s) (KPIs) [17, 12, 16, 18, 19].

In order to establish the relationship between the behavior at the resource level and the performance level in terms of SLA of the distributed system, the *Performance Forecaster* is in charge of collecting the low-level metrics (low-level resource use metrics) and high-level metrics (SLA parameters) in the profiling phase to generate a model capable of mapping them. This model will be the one used in auto-scaling phase to make the estimations of the high-level metrics that will be provided to the *Decider*.

Low-level metrics are provided to the *Performance Forecaster* through a monitoring service, which periodically collects utilization metrics. In the profiling phase, this service collects a wide variety of resource metrics, which after being pre-processed and transformed will be the predictors of the predictive performance model. In the auto-scaling phase, the model is able to predict the system's performance based on the values of the resource utilization metrics that the monitoring service delivers periodically. These performance estimations are sent to the *Decider* along with the *Workload Trend Forecaster* predictions.

*4.4. Decider*

The *Decider* is the module in charge of determining if it is necessary to trigger any scaling action. This decision is taken based on the information received from the previous modules, this is, the performance trend predicted in the prediction time horizon $h$ and the current performance of the system estimated in that moment. In addition, it needs some configuration parameters that establish the thresholds to decide which scaling action to perform, allowing flexibility and adaptation of this solution to different distributed systems. For the sake of clarity, this module is explained in more detail in Section 5.3, where the specific implementation of this module for a particular distributed system is explained.

# 5. FLAS for E-SilboPS

In recent years we have seen the increasing importance of publish-subscribe systems as a consequence of the strong adoption of event-driven architectures, where these systems are the cornerstone since they are in charge of sending the information asynchronously in the form of events [1, 13]. Compared to topic-based systems, content-based publish-subscribe systems allow subscribers to indicate their interests through predicates in a multi-dimensional system, which significantly reduces the processing of notifications by end users. To achieve this, they are usually implemented as distributed systems matching the incoming notifications to the stored subscriptions by determining which subscribers should receive each of the incoming notifications based on the interests described in each subscription.

This section describes how FLAS integrates with a high performance distributed system such as E-SilboPS. As mentioned, we have chosen to validate FLAS with E-SilboPS due to its greater complexity in its scaling actions that supports transparent, publisher-wise dynamic state repartitioning without client disconnection and with minimal notification delivery interruption for subscribers. This functionality makes the scaling time dependent on both the input workload and the current internal state of the system, which is a major challenge for the evaluation of an auto-scaling system such as FLAS since the estimation of the scaling time is quite variable.

More specifically, E-SilboPS was conceived by us as a content-based publish-subscribe middleware specially designed to be elastic [10, 21, 22, 23]. It is a distributed system composed by four layers of operators (Connection Point, Access Point, Matcher and Exit Point) forming a directed acyclic graph (DAG). Each of these operators can have a different number of instances, which can be increased or decreased independently by means of horizontal scaling operations (scale-out/in). The scaling algorithm allows dynamic distribution without disconnecting clients and with minimal interruption of the notification service.

It is important to emphasize that the prediction models are architecture agnostic, so it is a generic solution, since the implementation can differ as long as the defined API is respected. Nevertheless, the following sections describe the specific implementations that have shown good results for the case study exposed, as it can be seen from the analysis of the results of this evaluation.

### 5.1. Scaling Time Forecaster and Workload Trend Forecaster implementation

The *Scaling Time Forecaster* module is responsible for predicting the scaling time ($T'_{sa}$) based on the workload at a given instant, $T'_{sa}(workload)$. More specifically, as the load of content-based publish-subscribe systems is determined by the ratio of notifications per unit of time and stored subscriptions, $N$ and $S$ respectively, then the best possible function is pursued to calculate the time of a scaling action based on these parameters, $T'_{sa}(N, S)$. For the generation of the predictive model, in the profiling phase, the scaling times of several scaling actions with different workloads have been collected, in order to obtain a dataset that reflects the different scaling situations. Scaling actions during this phase are triggered in a reactive manner using different threshold-based rules on workload. With this training dataset, a linear regression model has been built that allows in the auto-scaling phase to determine the time of a scaling action based on the current workload.

The *Workload Trend Forecaster* is responsible for predicting the performance trend over a future prediction time horizon $h$. For the sake of simplicity and clarity, we have focused on response time as an SLA performance metric. In this case, the trend of performance is translated into the trend of response time expressed as the first order derivative of response time with respect to the time $\frac{\delta RT}{\delta t}$. In the profiling phase, time series of the response time are collected (top of Figure 3). A positive value of $\frac{\delta RT}{\delta t}$ will indicate an increasing trend and a negative value will indicate a decreasing trend with a more or less pronounced slope depending on the absolute value of the prediction. In this way, we do not directly forecast the response time or the workload, but rather the trend of the response time, which allows us to know how fast a response time that violates the SLA could be reached. To smooth this function and avoid fluctuations of the first order derivative, a Savitzky-Golay filter has been applied to the first order derivative of the data, which is a digital filter for smoothing the data, increasing its accuracy without distorting its trend (bottom of Figure 3). These smoothed data were used to generate various time series analysis models, after which the model with the lowest prediction error was chosen. This model is in charge of making the predictions on the trend of the response time in the future prediction time horizon $h$ in the auto-scaling phase. To generate these models, some time series analysis techniques have been evaluated, such as ARIMA, STL decomposition with an ETS model for the seasonally adjusted data and harmonic regression. In order to test these models, a cross-validation was performed with the models, choosing the one with the lowest prediction error (4).

During the auto-scaling phase, the *Workload Trend Forecaster* is in charge of forecasting in $t_0$ the values of $\frac{\delta RT}{\delta t_i}$ in the future instants $t_i = t_0 + T'_{sa}(t_0) + i$, $\forall i \in \{0, \dots h-1\}$ by means of the prediction model obtained in the profiling phase (Figure 5). In this way, we obtain a forecast of the response time trend in the $h$ future instants after finishing a possible scaling action that started at the current instant $t_0$. These forecast values are the ones that will be sent to the *Decider*. As already mentioned, the decision of which $h$-value (also known as forecast window size) to take is a complex one that depends on the degree of detail desired. On the one hand, a very small $h$-value can cause large fluctuations in the predicted values, while a too large value can omit fluctuations that are significant and should trigger a scaling action.

### 5.2. Performance Forecaster implementation

As previously mentioned, the monitoring service is in charge of collecting performance metrics in the monitoring phase and resource usage metrics in both the profiling and auto-scaling phases to send them to the *Performance Forecaster*. This service is executed periodically every second (the period is configurable) and collects a wide variety of resource usage metrics. For this implementation we have used the $dstat$[1] service, collecting more than 30 usage metrics of various resources such as processor (system, user, idle, wait, hardware interrupt, software interrupt, context switch metrics), memory (used, buffers, cache, free metrics), disk (read, write metrics), network (receive, send metrics), etc. Once these metrics are collected, the data is cleaned and pre-processed, adding compound metrics such as utilization percentages that will be reported to the *Performance Forecaster* in both execution phases.

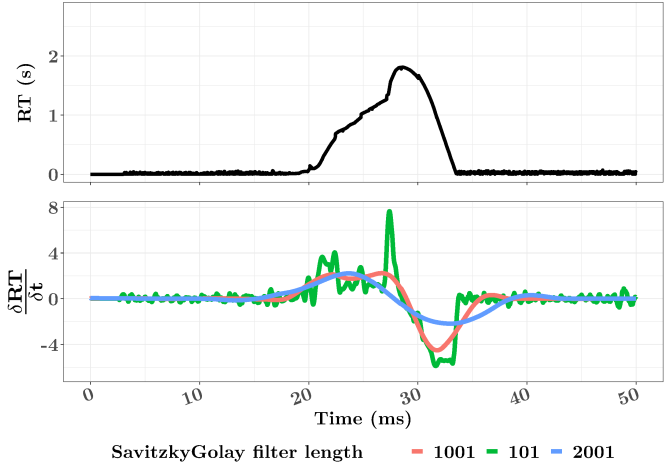

Figure 3: Part of a time series of the response time data before (top) and after applying a Savitzky-Golay digital filter for smoothing the data with different lengths (bottom).

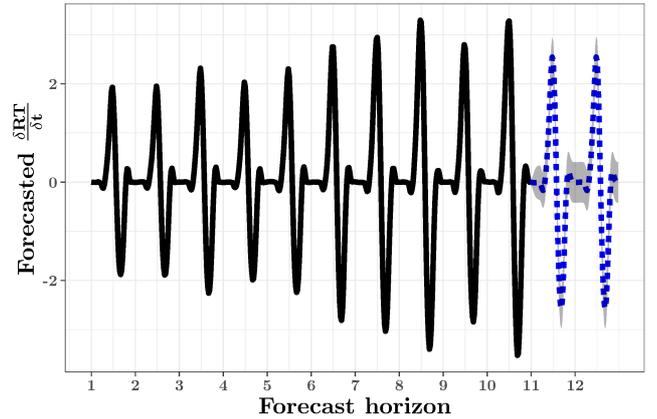

Figure 5: Forecasts of the values of $\frac{\delta RT}{\delta t_i}$ in a prediction time horizon using the harmonic regression model $ARIMA(2,0,2)$(dotted line). The horizontal axis represents the periods of seasonality.

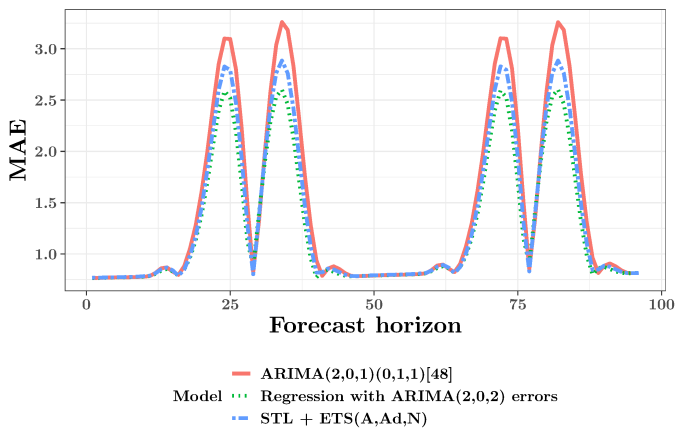

Figure 4: Comparison of the MAE value in the three time series analysis models tested by cross-validation for workload trend forecast.

The treatment of outliers is especially important, and that is why FLAS includes several mechanisms to treat them. In the cleaning and pre-processing phase of the data, the outliers are detected and removed. In addition, the values of the resource monitoring are average values of the sampling period. Finally, and as explained, FLAS scaling decisions require that appropriate conditions are maintained over time, and not at a single point.

A regression-based model has been chosen since statistical models, besides allowing us to make performance predictions in the auto-scaling phase, allow us to infer and understand the relationships between low and high-level metrics, being able to detect the KPIs of the application and the resources to be scaled in each scaling action.

More specifically, statistically valid linear regression models have been created for the two main performance metrics, throughput and response time. The results show that, as in the case of throughput, the low-level metrics that contribute the most information to the model (KPIs) are the amount of free RAM, the number context switches and the network usage (received and sent). On the other hand, the main KPI of the response time is the percentage of memory use, and the number of context changes, although to a lesser extent.

Once the mapping algorithms/models are chosen and trained for a given application type and for the KPIs considered in this paper, they remain static. They could need to be changed or trained again for a different set of KPIs (e.g associated to a different application type). They are dynamic in this sense, and this is why they are parameterizable in our system by design.

To exemplify this, we carried out a series of tests using four different publish-subscribe systems: E-SilboPS[10], RabbitMQ, ActiveMQ and Be-Tree[29] to analyze the type of relationship between low-level or resource-use metrics and high-level metrics or SLA parameters. From the results of these tests it was clear that content-based publish-subscribe applications (i.e. E-SilboPS and Be-Tree) were CPU bound, and in the case of topic-based publish-subscribe (RabbitMQ and ActiveMQ) they were memory bound.

---

[1]https://linux.die.net/man/1/dstat

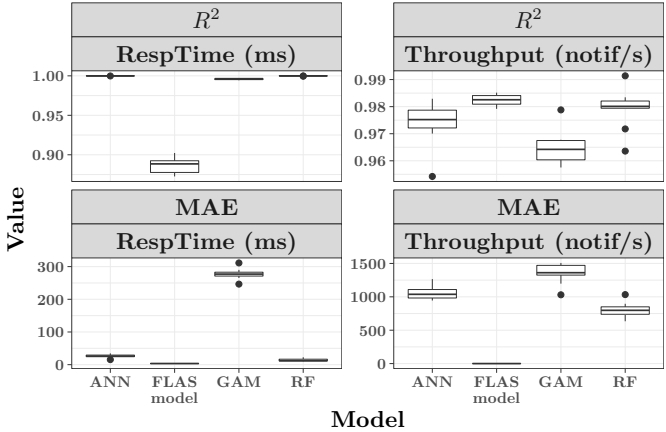

Figure 6: $R^2$ and $MAE$ values obtained from the 10-fold cross-validation performed with more than 40 predictive models of different types for comparison with the predictive models implemented in FLAS, both for response time and throughput (i.e. the model with the best results has been chosen as representative of its category, which is shown). The different categories of models are: Artificial Neural Networks (ANN), Generalized Additive Models (GAM) and Random Forests (RF).

Therefore, we can consider these relationships are static for the same application type (being it content-based or topic-based publish-subscribe), but not for different application types. Even more, we cannot ensure anything beyond that, not even within applications of the same paradigm, as is the case with publish-subscribe.

In order to evaluate the predictive capacity of these two models, a k-fold cross-validation (k=10) was performed with more than 40 predictive models generated for comparison of several types such as Random Forests (RF), Artificial Neural Networks (ANN), Generalized Additive Models (GAM) or Generalized Linear Models (GLM). In a very summarized way, due to the lack of space, Figure 6 shows the results of the 10-fold cross-validation ($R^2$ and $MAE$) comparing the FLAS predictive models for response time and for throughput with other types of predictive models (i.e. the model with the best results has been chosen as the representative model of each category).It can be seen how FLAS models have a very good predictive capacity ($R^2$) with a very low prediction error ($MAE$).

5.3. Decider

The *Decider* is the module in charge of gathering all the information from the previous modules to decide if any scaling action should be triggered, and if so, to decide the specific values of each of the scaling dimensions. This implementation of the *Decider* exploits the benefits of both approaches, predictive and reactive, since it initially checks the future predictions of the performance trend in order to take a decision in advance (proactive), but it also checks the current values of the estimated performance and compares them with some thresholds (reactive) as a contingency plan against possible failures of the predictive model.

The *Decider*, like the rest of modules in this implementation, is a service that runs periodically every second and executes the algorithm described in Algorithm 1. As can be seen, the main function of this module receives as parameters the current instant ($t_0$), the workload, a vector with the response time estimates in the moments prior to $t_0$ ($RT'$), and the Decider configuration, which contains a series of adjustable parameters of the *Decider* (i.e. $h$, $reactW$, $incTrendTH$, $decTrendTH$, $reactUpperTH$, $reactLowerTH$ and $majority$). The first verification it makes is that it is not currently in the cool-down time (line 1). After a scaling operation, a cool-down time is required to allow the system to stabilize and not trigger successive scaling actions continuously. The time of a possible scaling action ($T'_{sa}$) is then predicted (line 2) on the basis of the load (rate of incoming notifications per second and number of stored subscriptions). The response time trend prediction is obtained from the *Workload Trend Forecaster* by passing the current instant ($t_0$), the forecast scaling time ($T'_{sa}$) and the value of $h$ as an argument. As a result, a prediction vector is obtained of $\frac{\delta RT}{\delta t}$ for each of the time instants of $h$ (line 3). In addition, the resource usage metrics obtained from the *Monitoring Service* are passed as an argument to the *Performance Forecaster* to obtain an estimate of the response time at the current instant $RT'_{t_0}$ (lines 4 and 5) and this response time estimation is added to the vector $RT'$ (line 6) containing the previous RT forecasts.

Having made the corresponding predictions, the algorithm then decides whether any scaling action needs to be triggered. First, it checks if the predictions of $\frac{\delta RT}{\delta t}$ vector follow an upward trend (line 8). More specifically, the $incTrend()$ function checks that at least as many predictions of the time horizon $h$ as indicated by the $majority$ configuration parameter are above an $IncTrendTH$ defined also in the configuration. In this way, it can be verified whether, despite occasional fluctuations, the predicted values follow an increasing trend above a particular value. If the predictive condition of the scale-out is not fulfilled, the reactive condition is checked by calling the $RTAboveTH()$ function. This function checks whether the last $N$ response time estimations (reactive window, $reactW$ configuration parameter) are above a certain threshold ($reactUpperTH$) expressed in terms of the maximum response time specified by the SLA. If either of these two conditions are met, a scale-out action is triggered, doubling the number of Matcher instances and measuring the real time it takes to complete the scaling action, $T_{sa}$ (line 9). After any scaling action, the cool-down time is activated (calculated as a function of the $T_{sa}$ time previously measured) in which no scaling action can be performed (line 10). In addition, the response time estimation vector is cleaned so that the reactive condition can be reassessed (line 11). Similarly, the conditions for carrying out a scale-in action are assessed using the respective thresholds of the configuration (lines 15 to 20).

For the sake of clarity, some implementation details

**Algorithm 1:** *Decider* auto-scaling algorithm

**Input:** $t_0$, workload, RT', h, reactWindow, incTrendTH, decTrendTH, reactUpperTH, reactLowerTH, majority

1 **if** *coolDown == 0* **then**
2  $\quad T'_{sa} \leftarrow forecastT(workload.N, workload.S);$
3  $\quad \frac{\delta RT}{\delta t} \leftarrow forecastRTTrend(t_0, T'_{sa}, h);$
4  $\quad lowLevelMetrics \leftarrow monitoringService(t_0);$
5  $\quad RT'_{t_0} \leftarrow estimateRT(lowLevelMetrics);$
6  $\quad RT'.add(RT'_{t_0});$
7
    $\quad$// Scale-out evaluation
8  $\quad$**if** $incTrend(\frac{\delta RT}{\delta t}, incTrendTH, majoitry)$
    $\quad\quad || RTAboveTH(RT', reactUpperTH, reactW)$
    $\quad$**then**
9  $\quad\quad T_{sa} \leftarrow startScaleOut();$
10 $\quad\quad coolDown \leftarrow getCoolDownTime(T_{sa});$
11 $\quad\quad RT'.clear();$
12 $\quad\quad$**return**
13 $\quad$**end**
14
    $\quad$// Scale-in evaluation
15 $\quad$**if** $decTrend(\frac{\delta RT}{\delta t}, decTrendTH, majoitry)$
    $\quad\quad || RTBelowTH(RT', reactLowerTH, reactW)$
    $\quad$**then**
16 $\quad\quad T_{sa} \leftarrow startScaleIn();$
17 $\quad\quad coolDown \leftarrow getCoolDownTime(T_{sa});$
18 $\quad\quad RT'.clear();$
19 $\quad\quad$**return**
20 $\quad$**end**
21 **else**
22 $\quad$coolDown--;
23 **end**

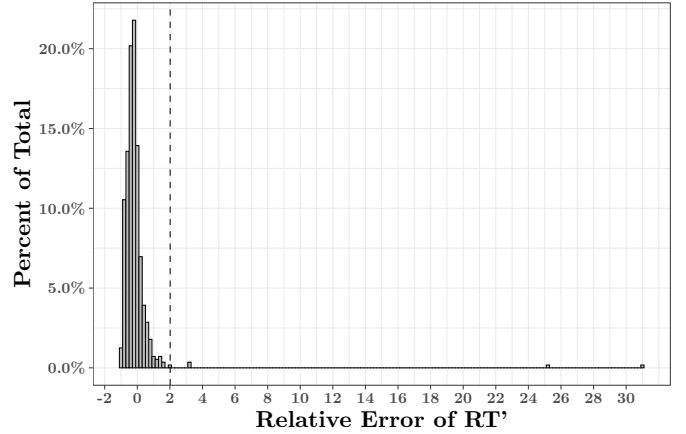

Figure 7: Relative frequency histogram and standardized density function of the relative error of response time estimation. The vast majority of the response time estimation error is constrained to low values (99 percentile shown by the vertical dashed line), and therefore the $RT'(t) \simeq RT(t)$ approach is used for reactive scaling.

such as synchronization between scaling operations have been omitted. This case reflects the horizontal scaling of the matchers (CPU-bound operator) which is the most complex case, since the scaling of the rest of operators is trivial as it has no state [10].

As seen, the reactive FLAS approach does not use the real response time (RT) metric, but the response time estimated by the *Performance Forecaster* (i.e. $RT'$). The approach $RT'(t) \simeq RT(t)$ for all $t$ instants allows that during the auto-scaling phase the application does not have to be monitored and therefore it is not necessary to instrumentalize the application in this phase, which makes FLAS a less invasive solution and reduces the monitoring overhead. However, it has been concluded that this approach is valid since the relative error of the response time estimation is limited to relatively low values for approximately 98% of the estimations (percentile 99), that is, values outside this range can be considered neither statistically frequent nor relevant, as shown in Figure 7. Although the relative error may seem high, it should be noted that the domain of response time estimation is very large (from values close to 0 to tens of thousands of milliseconds or more). In addition, the model makes its estimation based on a snapshot of the resource usage provided by the low-level metrics, which may cause that at a given moment there is a measurement of a punctual peak usage due to their large variability, which causes an estimation far above the real response time. However, it has been demonstrated that these cases are statistically not frequent and irrelevant (Figure 7).

## 6. Experimental evaluation

This section presents the evaluation of FLAS as an auto-scaling system for a distributed content-based publish-subscribe system such as E-SilboPS. In addition, the results of this evaluation are analyzed, showing how FLAS allows to minimize the time of violation of the performance SLA in different situations.

Unfortunately, there are no real public workloads available due to privacy concerns and commercial interests, which often hinders the validation of content-based publish-subscribe systems. Some works have been done as in [9], where the authors describe a possible solution, but it is not available for use. For this reason, several test cases with synthetic workloads have been generated for this evaluation. These test cases recreate several typical and expected functioning scenarios of the proposed solution (i.e. test cases 1 to 3). However, following the Boundary-Value Analysis (BVA) test methodology, other test cases are proposed that recreate some of the worst possible scenarios (i.e. test cases 4 and 5), thus complementing the evaluation made and showing how our solution works even in the worst possible cases, which are difficult to present in real contexts. The following subsections present and explain the different test cases proposed and the last subsection analyses the quantitative results obtained after multiple executions of the different test cases.

We ran our experiment on an Intel Core i7-4790K 4.00GHz and 16 GB of RAM running Linux 5.3.0-28-generic and openjdk 11.0.7. All test cases have been performed with an SLA that imposes a maximum response time of 1 second. Regarding the workload, the notifications are specially designed to match at least one subscription of those processed by E-SilboPS, which is a worst case scenario, since in a real scenario not all the notifications match at least one subscription. The publication rate of these notifications is constant in all test cases with a value of 10k notifications/s and the subscription dispatch rate is the one that varies throughout the test cases, since E-SilboPS is specially designed to scale in the number of subscriptions. This type of workload is typical of applications in contexts where the number of subscriptions is much higher than the number of notifications, such as stock market applications, or more recently, centralized applications for the notification and tracking of COVID-19 patients. Furthermore, all test cases start with a minimum configuration of 1-1-1 (1 Access/Connection Point, 1 Matcher and 1 Exit Point). In addition, the decider configuration parameters used in all the test cases have been the following:

- $h = 4$
- $incTrendTH = 0.9$
- $decTrendTH = -0.9$
- $reactUpperTH = 750ms$ (75% of the maximum RT allowed by the SLA)
- $reactLowerTH = 10ms$ (1% of the maximum RT allowed by the SLA)
- $reactW = 2$ (2 estimations must exceed the *ReactiveUpperThreshold* value to trigger a scaling action)

These specific values of the configuration parameters have been obtained empirically and have demonstrated the best results for this particular case. All the graphs shown in this section have been generated with these configuration parameters that have been empirically demonstrated to be the ones that produce the least number of SLA violations for this set of test cases. However, as can be seen in Section 6.6, other values of these parameters can generate different scaling decisions. More specifically, Section 6.6 compares the quantitative results of some of the alternative configurations tested and their comparison with the chosen configuration.

### 6.1. Test Case 1: Stationary peak

In this first test case there is a huge increase and later decrease in the number of subscriptions (up to 80k subs/s) in a time interval of a few seconds (peak), as shown by the dotted line in Figure 8. This peak occurs seasonally every certain period of time, representing a massive sending of subscriptions and subsequent unsubscriptions as a result, for example, of interest in an event that is repeated seasonally with the same period and that after some time loses interest (e.g. shopping channels with discount notifications before Christmas or the Black Friday, traffic news channels before holiday periods, etc). This situation, although infrequent, could cause a saturation that would degrade the service of other end users using the same E-SilboPS instance.

In this test case both scale-out and scale-in actions have been taken proactively due to the predictions of the *Workload Trend Forecaster*. As can be seen in Figure 8, the scale-out action is triggered before the response time grows exponentially, as a consequence of the forecast performance trend. Once the scale-out action is completed, it goes from a 1-1-1 topology to 1-2-1, doubling the number of Matcher instances, which doubles the processing capacity of the workload and reduces the response time. Later, when there is a forecast in which the trend is decreasing during $h$, the scale-in action is triggered, which reduces the Matcher instances to save resources, thus returning to a 1-1-1 topology. By means of the scale-out action, the maximum value of 1 second response time indicated by the SLA is not reached as in the case it would not have scaled (dashed line). Moreover, when the peak has passed and the processing demand is lower, the system releases the unnecessary resources by performing a scale-in action.

As shown in Figure 8, the time intervals between the vertical lines and the stroke and point lines indicates the forecast scaling time ($T'_{so}$ and $T'_{si}$ for the scale-out and scale-in respectively) and the time intervals between the two vertical lines and light shading indicates the actual time of the scaling action ($T_{so}$ and $T_{si}$). Both scaling actions end slightly earlier than predicted. In addition, due to the lack of a greater resolution cannot be seen how there is a slight increase in the response time when the scaling actions are initiated, this is because the scaling algorithm of the Matcheer in E-SilboPS requires to distribute the internal state between the total Matcher instances of the new configuration. This dynamic repartitioning requires sending the corresponding status to each instance and its processing, which implies an overhead to the processing of the workload that continues to be received during the scaling operations.

Finally, the dotted line represents the estimation of the response time in each one of those instants ($RT'$) that as can be seen is quite approximate to the value of the real response time measured in those instants ($RT$). A figure similar to this (Figure 8) can be observed in each of the following subsections where the test cases and their results are explained, and therefore, the way to interpret it is the same as explained here.

### 6.2. Test Case 2: Non-stationary peak

In this test case, there is a workload peak similar to the previous test case, but in a non-stationary manner. This change cancels out the predictive part of FLAS, since it is not able to predict when that peak will occur. However, FLAS is able to make the corresponding scaling decisions in a reactive manner to ensure compliance with the SLA. In a real context, this can occur if there is a programming error in the application using E-SilboPS, a denial of service (DoS) type attack, or simply a sudden and ephemeral massive interest in a certain entity.

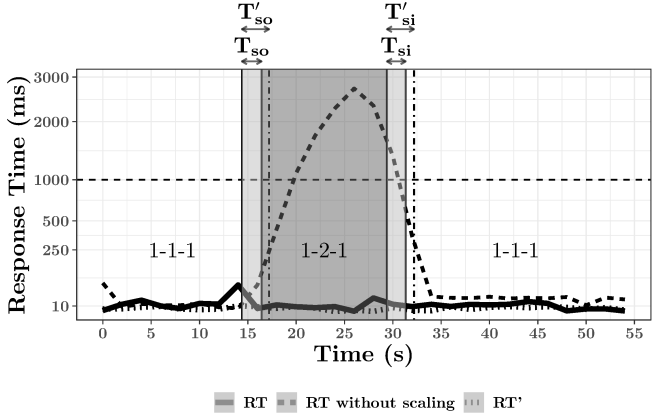
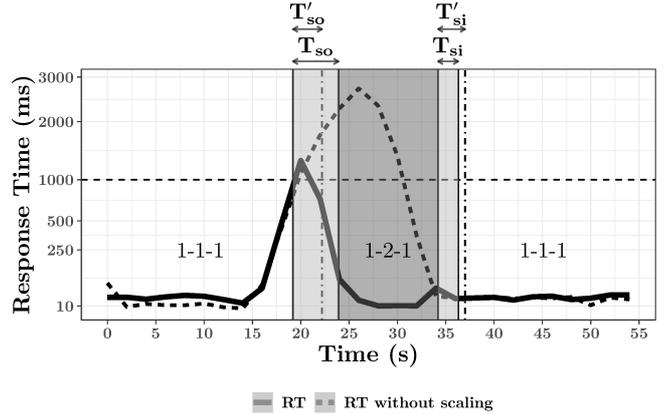

Figure 8: Test case of a stationary peak of subscriptions that FLAS is able to cope with by means of a predictive scale-out before the peak is reached and a scale-in also predictive when the workload is going to decrease permanently to save resources. The values of $T'_{so}$ and $T'_{si}$ indicate the forecast scaling time for the scale-out and scale-in actions respectively, while the values $T_{so}$ and $T_{si}$ indicate the actual times that these scaling actions have taken

Figure 9: Test case similar to that shown in Section 6.1 but using a reactive rather than predictive approach because it's non-stationary. A threshold of 75% of the maximum RT established by the SLA (i.e. 750 ms) has been used. Compared to Figure 8 it can be seen how the scaling action occurs later and therefore the response time increases.

Both scaling decisions (scale-out and scale-in) have been taken in a reactive manner using threshold based rules (Figure 9). When the RT predicted by the *Performance Forecaster* exceeds a certain threshold for a certain amount of time, the scaling action is triggered. Several executions have been performed with different values both of the threshold value (i.e. 75%, 80% and 90% of the maximum response time imposed by the SLA in the case of scale-out), and of the amount of time that should be exceeded to trigger the scaling action (i.e. reactive window size of 1, 2 and 3 consecutive estimations). These results have allowed us to empirically obtain the best FLAS configuration as detailed in Section 6.6 Quantitative analysis of results, where a quantitative comparison of all these variations can be seen.

In general, the reactive approach is less efficient than the predictive one, since it tends to incur longer periods of under-provision in the scale-out and over-provision in the scale-in as a result of its lack of anticipation. The higher the threshold value of the reactive scaling rules, the more these differences are accentuated. On the other hand, adding the restriction of exceeding the threshold value for more than one estimation prevents the triggering of scaling actions as a consequence of a false positive of the *Performance Forecaster*, but once again, it delays the decision making, and if it is not a false positive it could mean a higher SLA violation.

### 6.3. Test Case 3: Steady increase

This test case reflects a scenario in which subscriptions increase slowly but steadily over time (+500 subscriptions/s) until more than 100k subscriptions are stored. In a real context, this small but constant increase in load may correspond to a certain fashion or trend that causes subscriptions related to a certain entity to gradually increase as this entity becomes more popular, which means that large peaks are not reflected as in the previous cases (Figure 10). An example of this behavior can be social network profiles whose interest is slowly but steadily increasing over time or subscriptions to a platform that is gradually becoming popular.

This type of situation means that the slope predicted by the *Workload Trend Forecaster* is not sharp enough to trigger a scale-out in the first few moments, and when this slope is clearly accentuated, the maximum response time value imposed by the SLA is already being exceeded by a long margin. In this situation, a predictive scale-out is not possible because the scaling decision would be very late, once the SLA is violated which would boost even more the response time by adding the overhead of the scale-out process. Instead, the reactive approach we already discussed is able to react earlier in these cases, especially with low threshold values (i.e. 75% of the maximum response time set by the SLA). Moreover, even if the reactive approach determines a little late that a scale-out should be performed, in this context it is not so harmful because the increase of the response time is quite slow and therefore it will take longer to violate the maximum response time imposed by the SLA.

### 6.4. Test Case 4: Isolated and close to SLA limit workload peak

This test case is very similar to the first one, but unlike the first one, the peak of the workload occurs in a much shorter instant of time, going from a workload of 30k subscriptions/s to 120k subscriptions/s, to return the next instant to 30k subscriptions/s (Figure 11). As previously mentioned, this test case and next one represents a worst case scenarios following a BVA analysis, and therefore, it will be very difficult to find these type of case in a real context, but it allows to evaluate FLAS even in the worst imaginable conditions.

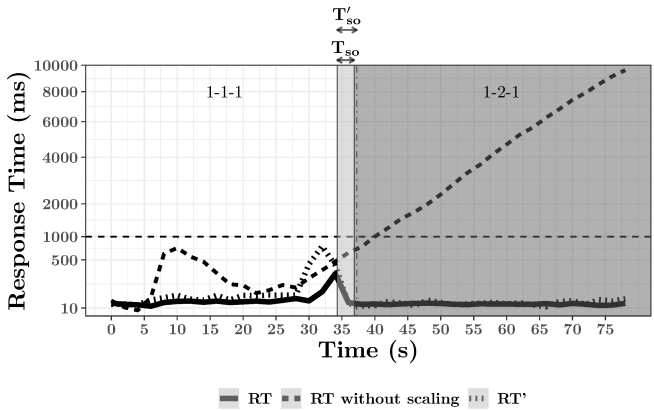
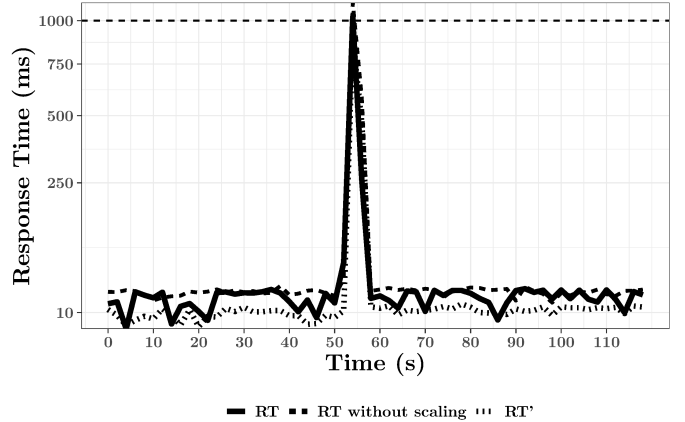

Figure 10: A test case in which the subscription rate increases slowly, which causes the response time to increase gradually and not generate a peak. In this case the predictive approach of FLAS is not able to predict sufficiently in advance the scale-out action, but the reactive approach allows FLAS to trigger the scale-out action without violating the SLA.

Figure 11: Test case with a very punctual peak workload in which FLAS determines that it is not efficient to trigger scaling actions (neither predictively nor reactively), therefore it has a 1-1-1 configuration during the whole test case.

Although this test case seems a little artificial, it serves to illustrate how the system is capable of adequately managing peaks so punctual that it is not worth triggering a scaling action, since the duration of the scale-out operation itself could be double or triple the duration of the peak, in addition the overhead introduced by the scaling action itself, could make the response time at that instant even greater. In this case, it is not efficient to trigger a scale-out and scale-in operation and FLAS is able to detect it through the condition that the trend has to be maintained during a certain number of the predictions of the prediction time horizon $h$. That is, the *Workload Trend Forecaster* detects the peak, but since it does not show a trend maintained over time, the *Decider* resolves that it should not trigger any scaling action. Moreover, in this case, the reactive window has to be large enough not to trigger a scale-out action as a result of the peak in response time estimation when the workload peak occurs. As shown in Figure 11, only the maximum response time SLA of one second is violated during a very short period of time (approximately 1 second).

*6.5. Test Case 5: Consecutive and close to SLA limit workload peaks*

This test case is based on the previous one, but instead of being a single punctual peak, several peaks occur in a row during a certain time interval. This behavior, as in the previous cases, is repeated periodically. In this case, it would be considered advisable to perform a scale-out action, since otherwise the maximum response time imposed by the SLA would be violated for quite some time as shown in Figure 12.

However, as in the previous test case, it is not possible to give a predictive response, as the upward trend is not stable over time, but fluctuates continuously. However, FLAS is still able to determine that it must perform a scale-out since it detects that the estimated response time is above a certain threshold (75% of the maximum established by the SLA, in this case, 750 ms) for two consecutive estimations. In this way, as shown in Figure 11, a reactive scale-out is produced. Some SLA violations can still occur in one or two seconds, but it is far less than if no scaling action were taken. Finally, scale-in is predictive, since it can detect a clear trend of decreasing response time once the succession of workload peaks is over.

*6.6. Quantitative analysis of results*

This section presents and discusses the results of the evaluation of the case studies described above (Tables 1 and 2). The results presented here are average values obtained from 20 executions of each test case. The default setting for the reactive scaling is the one mentioned above (75% of the maximum response time imposed by the SLA as a threshold for the scale-out and reactive window size of 2). It is later demonstrated that this is the most efficient configuration for this case.

When interpreting Tables 1 and 2, the distinction between NA and 0.00 values must be taken into account. NA values indicate that there is no data for that case, since it does not apply that measure for that particular case, while the other case indicates that the value of that metric is 0.00.

For each test case a single version has been executed without any scaling action in order to obtain a baseline of the response time behavior for that test case. This baseline has been used to determine the Demand Point of each scaling action ($DP_{sa}$). In each execution, the difference between a demand point for an scaling action and the completion of the corresponding scaling action ($RP_{sa}$), i.e. $DP_{sa} - RP_{sa}$, has been calculated. Positive values of this difference indicate over-provisioning in the case of a scale-out action and under-provisioning in the case of a scale-in action. On the other hand, negative values of this difference indicate under-provisioning and over-provisioning if the scaling action is scale-out and scale-in respectively. In this way, the percentage of time in over-provisioning and under-provisioning can be calculated as the sum of all over-provisioning and under-provisioning times for each scaling action divided by the total runtime.

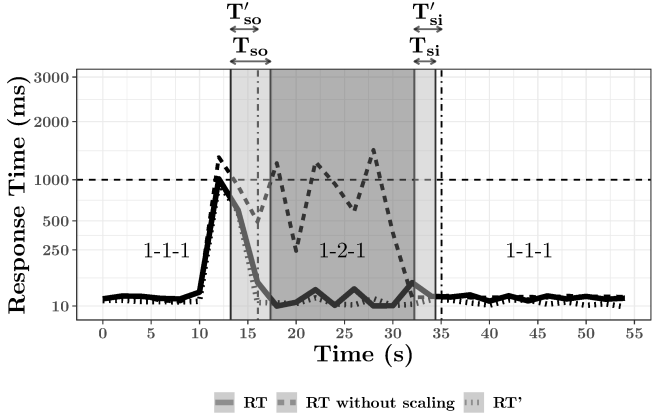

Figure 12: Test case with continuous punctual load peaks. FLAS is able to perform a scale-out to reactively to cope with the increased workload. The scale-in is predictive, since it predicts a continuous decrease of the response time. Only one estimation above the reactive threshold can be appreciated but this is due to the aggregation of the data for the sake of clarity.

Finally, Table 1 also shows the percentage of time that the SLA has been violated, that is, the percentage of time that the $RT$ has been greater than 1 second. Table 2 shows the scale action times calculated as $\sum |TP_{sa} - RP_{sa}|$ for all scale-in and scale-out actions respectively. In addition, for each type of scaling action it shows the relative error of the prediction of $T$, calculated as $\sum \frac{T_{forecast} - T}{T}$. The sign of this relative prediction error indicates if the predicted $T$ value is greater than the real value of the scaling time (positive) or if, on the contrary, the predicted scaling time is less than the real time of the scaling action (negative).

The Test case 5 has a very low SLA maximum response time violation rate (0.36%). The over-provisioning of this test case is mainly due to the predictive scale-in actions and the under-provisioning to the reactive scale-out actions, being the latter the ones that cause these small SLA violations. In turn, the Test case 3 does not violate the maximum response time of the SLA at any time due to the early reactive scaling as seen in the percentage of over-provisioning time, the slow increase in response time of this test case and the null value of under-provisioning.

The Test case 4 is the only case where no scaling action is taken and yet the SLA violation percentage is very low (less than 1%). It should be remembered that in this case, since the peak is so small, it is not profitable to perform a scale-out, since the overhead of the scale-out action would produce an increase in response time that would result in a larger SLA violation than if this scale-out action is not performed. Due to the reactive scaling window, FLAS is able to detect this situation and conclude that it is not necessary to trigger the scale-out action.

For the case of the Non-stationary peak test (Test case 2) several reactive versions have been test to be able to be compared with the predictive one (Test case 1) and show how the configuration values of the *Decider* have been fixed empirically. Each version of Test case 2 is identified as X-Y, where X is the scale-out threshold (as a percentage of the maximum RT defined by the SLA), and Y is the size of the reactive window. Therefore, the higher the X and Y values, the later the scaling action will be triggered. As can be seen, the predictive version (Test case 1) does not violate the SLA at any time, since the under-provisioning is very low and the over-provisioning is somewhat higher (due to the prediction error). Looking at the reactive versions, for the same value of the scaling threshold, the total under-provisioning time increases as the size of the reactive scaling window increases, and therefore, the time in violation of the SLA increases. In view of this data, the reactive configuration 75-1 and 75-2 are the ones that generate less SLA violations, and specifically, the implementation of configuration 75-2 has been chosen in the *Decider* since it allows to efficiently treat cases such as Test case 4, while with configuration 75-1 it would not be possible (although with configuration 75-2 SLA violations slightly increase compared to 75-1, they still do not exceed 1% of the time).

With respect to Table 2, it can be seen how the scale-out time (*avg T scale-out*) increases with the size of the reactive window and the scale-out threshold. This is due to the fact that, as mentioned, with higher values of the reactive window and the scale-out threshold, the scale-out action is triggered later, and therefore the E-SilboPS load is higher and in saturation, which increases the scale-out time. This same phenomenon can be observed in the relative error of prediction of T in the scale-out (last column), since it is observed that for reactive cases, the error goes from positive values (i.e. $T' > T$), to negative values (i.e. $T' < T$).

In Table 2 it can be seen how for the scaling actions triggered in a predictive way, the relative error of prediction of $T$ oscillates between approximately 30% and 40%. This *a priori* could be translated in a bad accuracy of the predictive model of the *Scaling Time Forecaster*, but it is necessary to analyze it in this context. The scaling horizon $h$ of the *Workload Trend Forecaster* depended on the prediction of $T$, so if the predicted $T$ is always slightly higher, this causes that $h$ is also larger, and therefore the scaling decisions are taken slightly in advance of what would be the optimum. This leads to a slight increase in the percentage of time in over-provisioning in those same cases, as can be seen in Table 1. However, it must be taken into account that in this context the objective of FLAS is to minimize performance SLA violations, and therefore, that the scaling actions are slightly advanced (as a consequence of a predicted T greater than the real $T$) entails a higher cost derived from a higher over-provisioning. However, it should be borne in mind that a slightly delayed scaling action (predicted $T$ less than actual $T$) usually entails an increase in SLA violations (as seen in the results of Table 1), which is a much greater penalty than a slight over-provisioning, and therefore, in this context it seems cautious to have that margin of error.

| Test case | Reactive Window Size | Total over-provisioning time (%) | Total under-provisioning time (%) | Time in SLA violation (%) |
|---|---|---|---|---|
| Test case 5 | 2 | 10.16 | 7.39 | 0.36 |
| Test case 3 | 2 | 3.98 | 0.00 | 0.00 |
| Test case 4 | 2 | NA | NA | 0.61 |
| Test case 1 | 2 | 7.66 | 0.84 | 0.00 |
| Test case 2 ($reactUpperTH = 750ms$) | 1 | 7.65 | 0.84 | 0.18 |
| | 2 | 7.61 | 2.16 | 0.78 |
| | 3 | 9.18 | 8.13 | 6.14 |
| Test case 2 ($reactUpperTH = 800ms$) | 1 | 8.87 | 3.20 | 1.56 |
| | 2 | 7.02 | 4.92 | 5.21 |
| | 3 | 12.10 | 11.02 | 5.71 |
| Test case 2 ($reactUpperTH = 900ms$) | 1 | 9.00 | 4.72 | 2.27 |
| | 2 | 8.08 | 4.45 | 2.73 |
| | 3 | 10.07 | 9.54 | 7.79 |

Table 1: Results of the percentage of under-provisioning, over-provisioning and violation of the maximum response time imposed by the SLA for the different test cases. In the case of the Stationary Peak test case, the predictive and reactive X-Y versions are compared (being X the scale-out threshold and Y the size of the reactive window respectively).

| Test case | Reactive Window Size | avg T scale-in (s) | avg T scale-out (s) | Relative error of T prediction in scale-in (%) | Relative error of T prediction in scale-out (%) |
|---|---|---|---|---|---|
| Test case 5 | 2 | 1.99 | 2.88 | 40.71 | -2.84 |
| Test case 3 | 2 | NA | 2.93 | NA | 2.86 |
| Test case 4 | 2 | NA | NA | NA | NA |
| Test case 1 | 2 | 1.97 | 2.11 | 42.31 | 36.81 |
| Test case 2 ($reactUpperTH = 750ms$) | 1 | 1.99 | 2.92 | 41.05 | 5.31 |
| | 2 | 2.29 | 3.19 | 28.87 | -0.57 |
| | 3 | 2.02 | 5.39 | 38.51 | -41.03 |
| Test case 2 ($reactUpperTH = 800ms$) | 1 | 2.06 | 3.27 | 36.07 | 0.11 |
| | 2 | 1.41 | 4.32 | 99.07 | -10.40 |
| | 3 | 1.99 | 5.33 | 41.13 | -38.50 |
| Test case 2 ($reactUpperTH = 900ms$) | 1 | 1.99 | 3.79 | 41.07 | -12.70 |
| | 2 | 1.97 | 3.89 | 41.91 | -16.91 |
| | 3 | 2.14 | 5.80 | 33.12 | -43.81 |

Table 2: Results of the scaling times ($T$) and the error in the prediction of these times for the scale-out and scale-in actions of the different test cases.

Following the evaluation methodology of similar works[17], Figure 13 compares FLAS with each one of auto-scaling techniques that compose it (reactive and predictive) and with a reactive technique using threshold-based rules. The latter has been chosen because it is the most widely used auto-scaler in practice, since it is the one offered by the main Cloud providers. In this case, it has been configured so that the scale-out is triggered if CPU usage is above 80% during more than two monitoring periods, and the scale-in is triggered in an analogous way with a threshold of 40%. This configuration has been chosen because it has been empirically proven to minimize SLA violations. After any scaling action of any of the auto-scalers, the same cool down time has been used.

As can be seen in Figure 13, the percentage of time in violation of FLAS SLAs is the lowest of the two techniques that compose it separately, that is, FLAS is capable of coordinating the two auto-scaler techniques to choose the most favorable one in any situation, even in the most extreme cases represented here. On the other hand, it can also be seen how the reactive technique that uses threshold-based rules has a percentage of SLA violation very similar to that of FLAS, and therefore better than the auto-scaling techniques of FLAS separately in some test cases. However, it achieves this with an over-provisioning much higher than FLAS and its separate parts in all test cases, which results in an increase of the economic cost and energy consumption compared to FLAS.

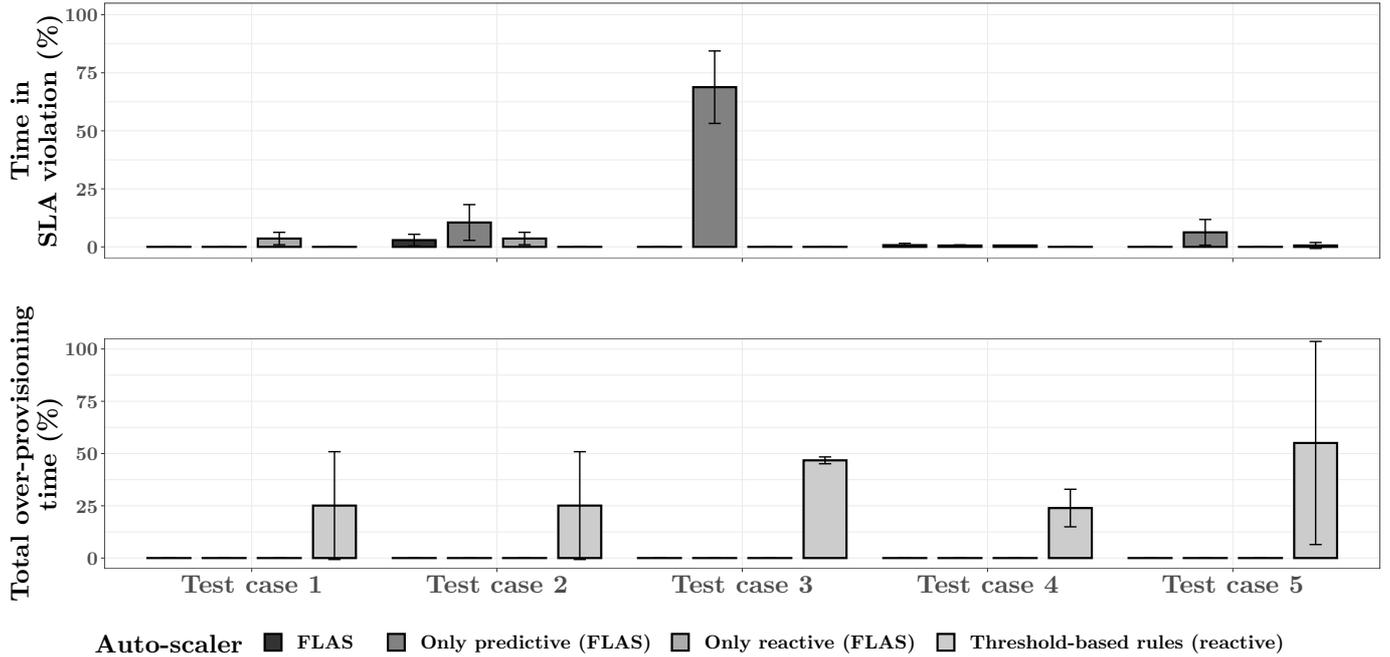

Figure 13: Comparison of FLAS with other auto-scaling techniques in terms of percentage of time that the SLA is violated and percentage of time in over-provisioning.

Finally, considering these results we can conclude that FLAS meets its objective of minimizing performance SLA violations, especially since some of the test cases presented here are the worst possible scenarios. In these circumstances (taking into account that reactive versions with configuration different from 75-2 are shown only for comparison), it is verified that the percentage of time in which the maximum response time imposed by the SLA of 1 second is violated is less than 1% (i.e. 0.78% in the worst case), which ensures a compliance with the SLA of 99.22% of the time. Furthermore, it has been demonstrated that FLAS has a better performance than each of the auto-scaling techniques that compose it, being able to coordinate both to choose the most convenient one in each situation. When compared to other auto-scalers, FLAS has demonstrated to obtain similar results in terms of SLA compliance, but with a quite inferior over-provisioning, which results in a lower economic cost and energy savings using FLAS.

## 7. Conclusions

In this paper we have presented FLAS (Forecasted Load Auto-Scaling) a generic distributed system auto-scaler by combining the advantages of both auto-scaling approaches, proactive and reactive. The main contributions of this paper are (i) the design and definition of the architecture of a generic framework for distributed system auto-scaling through the combination of proactive and reactive techniques, and (ii) a solution for the auto-scaling of a content-based publish-subscribe distributed system (E-SilboPS). The problems that have been addressed by implementing the architectural framework for the case study have been (i) to create a model to predict the trend of relevant SLA parameters (i.e. response time or throughput), (ii) to create a model capable of estimating, agnostically with respect to the type of application, SLA parameters based on resource usage metrics, (iii) to create a model to predict the scaling time and (iv) a mainly proactive auto-scaling algorithm that also contemplates a reactive scaling as a contingency to possible prediction failures. In addition, the modular and decoupled design of the FLAS architecture allows it to adapt to different distributed systems by modifying few configuration parameters and to replace the predictive models of the different modules of its architecture while respecting their definition.

Experimental results show how FLAS is able to meet the performance requirements set by the SLA even in the worst possible cases. More specifically, it has been demonstrated how FLAS enables a content-based publish-subscribe system to meet the maximum response time imposed by the SLA for over 99% of the time by taking the necessary scaling decisions at any given time under a wide variety of workloads by exploiting the advantages of both reactive and proactive approaches. In addition, it has also been demonstrated how its scaling algorithm is able to determine those cases in which triggering an scaling action is not efficient, since the overhead involved in the scaling action could lead to a violation of the SLA greater than that incurred in the case of not scaling. To the best of our knowledge, this would be the first auto-scaler for content-based publish-subscribe systems.

Furthermore, it has been empirically demonstrated how FLAS efficiently coordinates both the reactive and predictive auto-scaling approaches that compose it to offer better or equal SLA compliance than each of them separately. Compared to other auto-scalers, FLAS achieves the same positive results as other auto-scalers in compliance with SLAs but with a much lower over-provisioning, reducing the economic cost and the energy consumption.

## 8. Future directions

This work addresses all dimensions of scaling, however, the current implementation of FLAS duplicates or divides by two the resources to be added or removed as a first approach to the dimension of how much to scale. We are currently working on a module that allows predicting the minimum configuration required for the performance predictions obtained. However, this is a module whose configuration is quite specific to the application and therefore not as generic for all distributed systems as the rest of the FLAS architecture that we have presented in this work. For example, the E-SilboPS with which we have evaluated FLAS in this work allows to deploy or remove several instances of an operator in a single scaling operation, but other systems do not allow this option and must sequence successive scaling operations.

In addition, we are currently working to improve workload trend prediction in order to predict non-stationary workloads.

Currently, the prediction of the time of a scaling action ($T'_{sa}$) is determined by the load, more specifically, in the case of E-SilboPS we have seen that it is based on the ratio of notifications/s and the load of processed subscriptions. We believe that it would be interesting to add as a predictor to this model the prediction of the load in future moments, to determine more precisely the time of the scaling action and avoid possible uncontrolled peaks of response time when the scaling algorithm of the application involves an overhead (dynamic distribution of the state of the operators).

Finally, other improvements that are being studied are, on the one hand, the inclusion of the current configuration of the distributed system as an additional predictor of the predictive models to test whether the configuration of a distributed system at a given time can condition future predictions. On the other hand, now that the relationships between low-level and high-level metrics have been studied by means of a statistical predictive model, another predictive model could be developed with other methods (i.e. Artificial Neural Networks) that take into account these discovered relationships. However, although more complex predictive techniques can improve the accuracy of the predictive approach of FLAS, it should be taken into account that this can impact on the performance of FLAS causing a considerable overhead.